 \documentclass[aps,prd,preprint,preprintnumbers,showpacs,showkeys,nofootinbib, superscriptaddress,floatfix]{revtex4-1}
\usepackage{amsmath,amssymb,bm,mathrsfs}
\usepackage[colorlinks=true,citecolor=blue,linkcolor=blue]{hyperref}

\begin{document}
\title{Kinetic Term Anarchy for Polynomial Chaotic Inflation}

\author{Kazunori Nakayama}
\email{kazunori@hep-th.phys.s.u-tokyo.ac.jp}
\affiliation{Department of Physics, University of Tokyo, Tokyo 113-0033, Japan}
\affiliation{Kavli Institute for the Physics and Mathematics of the
  Universe (WPI), Todai Institutes for Advanced Study, University of Tokyo,
  Kashiwa 277-8583, Japan}

\author{Fuminobu Takahashi}
\email{fumi@tuhep.phys.tohoku.ac.jp}
\affiliation{Department of Physics, Tohoku University, Sendai 980-8578, Japan}
\affiliation{Kavli Institute for the Physics and Mathematics of the
  Universe (WPI), Todai Institutes for Advanced Study, University of Tokyo,
  Kashiwa 277-8583, Japan}

\author{Tsutomu T. Yanagida}
\email{tsutomu.tyanagida@ipmu.jp}
\affiliation{Kavli Institute for the Physics and Mathematics of the
  Universe (WPI), Todai Institutes for Advanced Study, University of Tokyo,
  Kashiwa 277-8583, Japan}

\begin{abstract}
We argue that there may arise a relatively flat inflaton potential over super-Planckian field 
values with an approximate shift symmetry, if the coefficients of the kinetic terms for many singlet 
scalars are subject to a certain random distribution. The inflaton potential 
generically contains various shift-symmetry breaking terms, leading to a possibly large deviation of the predicted values of the spectral index and
tensor-to-scalar ratio from those of the simple quadratic chaotic inflation. We revisit a polynomial chaotic 
inflation in supergravity as such. 
\end{abstract}
\preprint{UT-14-30, TU-973, IPMU14-0142}
\maketitle

%
%
\def\a{\alpha}
\def\b{\beta}
\def\c{\varepsilon}
\def\d{\delta}
\def\e{\epsilon}
\def\f{\phi}
\def\g{\gamma}
\def\h{\theta}
\def\k{\kappa}
\def\l{\lambda}
\def\m{\mu}
\def\n{\nu}
\def\p{\psi}
\def\q{\partial}
\def\r{\rho}
\def\s{\sigma}
\def\t{\tau}
\def\u{\upsilon}
\def\v{\varphi}
\def\w{\omega}
\def\x{\xi}
\def\y{\eta}
\def\z{\zeta}
\def\D{\Delta}
\def\G{\Gamma}
\def\H{\Theta}
\def\L{\Lambda}
\def\F{\Phi}
\def\P{\Psi}
\def\S{\Sigma}

\def\o{\over}
\def\beq{\begin{eqnarray}}
\def\eeq{\end{eqnarray}}
\newcommand{\gsim}{ \mathop{}_{\textstyle \sim}^{\textstyle >} }
\newcommand{\lsim}{ \mathop{}_{\textstyle \sim}^{\textstyle <} }
\newcommand{\vev}[1]{ \left\langle {#1} \right\rangle }
\newcommand{\bra}[1]{ \langle {#1} | }
\newcommand{\ket}[1]{ | {#1} \rangle }
\newcommand{\EV}{ {\rm eV} }
\newcommand{\KEV}{ {\rm keV} }
\newcommand{\MEV}{ {\rm MeV} }
\newcommand{\GEV}{ {\rm GeV} }
\newcommand{\TEV}{ {\rm TeV} }
\def\diag{\mathop{\rm diag}\nolimits}
\def\Spin{\mathop{\rm Spin}}
\def\SO{\mathop{\rm SO}}
\def\O{\mathop{\rm O}}
\def\SU{\mathop{\rm SU}}
\def\U{\mathop{\rm U}}
\def\Sp{\mathop{\rm Sp}}
\def\SL{\mathop{\rm SL}}
\def\tr{\mathop{\rm tr}}

\def\IJMP{Int.~J.~Mod.~Phys. }
\def\MPL{Mod.~Phys.~Lett. }
\def\NP{Nucl.~Phys. }
\def\PL{Phys.~Lett. }
\def\PR{Phys.~Rev. }
\def\PRL{Phys.~Rev.~Lett. }
\def\PTP{Prog.~Theor.~Phys. }
\def\ZP{Z.~Phys. }

\newcommand{\bea}{\begin{eqnarray}}   
\newcommand{\eea}{\end{eqnarray}}
\newcommand{\bear}{\begin{array}}  
\newcommand {\eear}{\end{array}}
\newcommand{\bef}{\begin{figure}}  
\newcommand {\eef}{\end{figure}}
\newcommand{\bec}{\begin{center}}  
\newcommand {\eec}{\end{center}}
\newcommand{\non}{\nonumber}  
\newcommand {\eqn}[1]{\beq {#1}\eeq}
\newcommand{\la}{\left\langle}  
\newcommand{\ra}{\right\rangle}
\newcommand{\ds}{\displaystyle}
\def\SEC#1{Sec.~\ref{#1}}
\def\FIG#1{Fig.~\ref{#1}}
\def\EQ#1{Eq.~(\ref{#1})}
\def\EQS#1{Eqs.~(\ref{#1})}
\def\TEV#1{10^{#1}{\rm\,TeV}}
\def\GEV#1{10^{#1}{\rm\,GeV}}
\def\MEV#1{10^{#1}{\rm\,MeV}}
\def\KEV#1{10^{#1}{\rm\,keV}}
\def\lrf#1#2{ \left(\frac{#1}{#2}\right)}
\def\lrfp#1#2#3{ \left(\frac{#1}{#2} \right)^{#3}}
\def\REF#1{Ref.~\cite{#1}}
\newcommand{\osc}{{\rm osc}}
\newcommand{\ed}{{\rm end}}
\def\dda#1{\frac{\partial}{\partial a_{#1}}}
\def\ddat#1{\frac{\partial^2}{\partial a_{#1}^2}}
\def\dd#1#2{\frac{\partial #1}{\partial #2}}
\def\ddt#1#2{\frac{\partial^2 #1}{\partial #2^2}}
\def\lrp#1#2{\left( #1 \right)^{#2}}
%
%

\section{Introduction}
Inflation elegantly solves various problems of the standard big bang cosmology such
as the horizon and flatness problems \cite{Guth:1980zm,Kazanas:1980tx}.
The slow-roll inflation paradigm~\cite{Linde:1981mu, Albrecht:1982wi} has enjoyed a remarkable success in explaining observations of cosmic microwave background  
and large-scale structure. Measuring the primordial B-mode polarization will be a further important step
toward a proof of inflation.\footnote{Recently the BICEP2 experiment~\cite{Ade:2014xna}
 announced  detection of the primordial B-mode polarization.
Other observations such as Planck are necessary to confirm/refute the result. }

There are two important parameters, the spectral index, $n_s$, and the tensor-to-scalar ratio, $r$,
which characterize the metric perturbations generated during inflation.  
The on-going and planned B-mode experiments are targeting $r \gtrsim 10^{-2} - 10^{-3}$, for which
the inflaton field excursion during the last $50$ or $60$ e-foldings exceeds the Planck scale~\cite{Lyth:1996im}.
For  inflation models with $r$ within the expected sensitivity of the foreseeable experiments, 
therefore, one has to have a good control of the inflaton potential over super-Planckian field ranges.
This places a stringent constraint on the inflation model building. 
The problem becomes acute in  supergravity, where the scalar potential exponentially blows up
 beyond the Planck scale for a general form of the K\"ahler potential.
 
In the pioneering paper by Kawasaki, Yamaguchi and one of the present authors (TTY)~\cite{Kawasaki:2000yn,Kawasaki:2000ws},  they proposed a simple prescription to realize a chaotic inflation model~\cite{Linde:1983gd} in supergravity by introducing a shift symmetry
\bea
\Phi \to \Phi + i \alpha,
\eea
where $\alpha$ is a real transformation parameter. The K\"ahler potential is assumed to respect the above
shift symmetry, but the superpotential contains explicit breaking terms:
\bea
K&=& \frac{1}{2}(\Phi+\Phi^\dag)^2 + |X|^2 + \cdots,\\
W &=& M X \Phi,
\label{WKYY}
\eea
where $\Phi$ forms a Dirac mass term with another chiral superfield $X$.  The imaginary component of $\Phi$ has a quadratic 
potential up to super-Planckian field values thanks to the shift symmetry. There are many  variants and extensions of the chaotic inflation model in the same spirit. See e.g. Refs.~\cite{Takahashi:2010ky,Nakayama:2010kt,Kallosh:2010ug,Kallosh:2010xz,Harigaya:2012pg,Nakayama:2013jka,Nakayama:2013txa,Harigaya:2014qza,Kallosh:2014xwa}. 
The inflaton potential, and therefore the predicted values of the spectral index $n_s$ and tensor-to-scalar
ratio $r$ sensitively depend on the nature of the shift-symmetry breaking terms.

One of the central issues in large-field inflation models is the quality of the  shift symmetry.
If the symmetry is of high quality, the inflaton potential will be relatively flat and remain under control 
up to field values much greater than the Planck scale. On the other hand, if the shift symmetry is 
only approximate (or accidental),  the inflaton potential might become extremely steep for the 
field values greater than a few ten times the Planck scale. In the latter case, the flatness of the inflaton potential, or
the quality of the shift symmetry, might be due to the anthropic requirement 
that our universe experience inflation with the total e-folding number $50$ or $60$.

In this letter we argue that the shift symmetry required for successful large-field inflation may arise
accidentally as a consequence of random-valued coefficients of the kinetic terms for 
many singlets, to which we refer as ``kinetic term anarchy", 
by analogy of the neutrino mixing anarchy hypothesis~\cite{Hall:1999sn,Haba:2000be}.  
That is to say, if  the kinetic term coefficient for one of the singlets becomes  much larger than unity,
the corresponding scalar potential becomes relatively flat over super-Planckian field ranges. This may happen especially if there are many singlets. In supergravity, the large kinetic term
coefficient is not enough to realize large-field inflation, and one needs a certain relation
between holomorphic and non-holomorphic quadratic terms. We shall consider two cases as to the
origin of such relation.

Lastly, let us here mention related works in the past. 
It was noted in Refs.~\cite{Dimopoulos:2003iy,Izawa:2007qa} that a large wave function factor helps to realize the 
slow-roll inflation. The origin of the large factor could be due to a modulus field~\cite{Dimopoulos:2003iy} or
the inflaton itself~\cite{Takahashi:2010ky,Nakayama:2010kt}. The key difference of the present work 
from these works is that we introduce many singlet scalars with random-valued kinetic term coefficients in order
to realize the large factor. Interestingly, when applied to the right-handed sneutrino
chaotic inflation~\cite{Murayama:1992ua,Nakayama:2013nya,Murayama:2014saa},
such anarchic nature may partly explain the observed large neutrino mixing angles.

\section{Kinetic term anarchy for large-field inflation}  \label{sec:kin}
In order to see how the shift symmetry for the successful inflation could arise, let us consider two toy models.
In the first model, we introduce many singlet scalars, which respect the shift symmetry for the quadratic terms
in the K\"ahler potential, whereas it is explicitly broken by various Planck-suppressed operators in the K\"ahler and super-potentials.
In the second model, we discuss a possibility that the shift symmetry considered in the first one arises accidentally.
Throughout this letter  we adopt the Planck units in which  the reduced Planck mass $M_P \simeq 2.4 \times \GEV{18}$ 
is set to be unity.  

\subsection{Model with shift symmetry at the renormalizable level}
First let us introduce $N$ singlet chiral superfield $\phi_i$  $(i=1-N)$ with the following K\"ahler potential:
\bea
K &=& \frac{1}{2} Z_{i j} (\phi_i + \phi_i^\dag) (\phi_j+\phi_j^\dag)   + \cdots,
\label{K1}
\eea
where $Z_{ij}$ is a real symmetric matrix representing the kinetic term coefficients, 
and the sum over repeated indices is understood.
We have assumed
that the scalars respect the shift symmetry,\footnote{
With an abuse of notation, we shall use the same symbol to denote both a chiral superfield and its lowest component.
} 
\beq
\phi_i \to \phi_i + i \alpha
\eeq
at the renormalizable level. 
Here and in what follows we assume $\phi_i$ has vanishing R-charges and we impose a $Z_2$ symmetry $\phi_i \to - \phi_i$, although most of the following argument can be  straightforwardly applied to the case without the $Z_2$ symmetry. Note that, because of the shift symmetry, the imaginary components of $\{\phi_i\}$ do not appear in the quadratic terms.

The shift symmetry is assumed to be explicitly broken by  various
Planck-suppressed operators such as $K \supset \phi_i \phi_j \phi_k \phi_\ell$, $\phi_i \phi_j \phi_k^\dag \phi_\ell^\dag$, etc., which are collectively represented by the dots in Eq.~(\ref{K1}). Specifically we assume that 
those Planck-suppressed operators are such that each ${\rm Im}[\phi_i]$ can take values up to the 
Planck scale without exponential steepening of the potential,  when the other scalars are set to be at the origin. 
This condition on the higher dimensional operators do not depend on the quadratic term coefficient $Z_{ij}$, 
because the imaginary components of $\{\phi_i\}$ appear only in the higher dimensional operators 
owing to the shift symmetry. 

Our main idea is as follows. Let us diagonalize the kinetic term coefficients $Z_{ij}$ by an orthogonal 
matrix:
\bea
\label{phi}
  { \phi} &=& R {\tilde \phi},\\
  \label{RZR}
R^T Z  R  &=& {\rm diag}(z_i) \equiv D_z
\eea
with
\bea
0 < z_1 \leq z_2\leq  \cdots \leq z_N, 
\eea
 where $R$ is an $N \times N$ orthogonal matrix, and we have suppressed the indices in (\ref{phi}) and
 (\ref{RZR}).  The canonically normalized fields $\{ \Phi_i\}$ are given by
 \beq
\Phi =  \sqrt{D_z} R^T \phi.
\label{Phi}
 \eeq
If the kinetic term coefficient matrix $Z_{ij}$ is subject to a certain random distribution, 
its largest eigenvalue, $z_N$,  could be much larger than the typical eigenvalues
which we set to be of order unity.
It implies that, 
in terms of the canonically normalized field, $\Phi_N = \sqrt{z_N} \tilde \phi_N$, its imaginary
component can take super-Planckian field values, $|{\rm Im}[\Phi_N]| \lesssim \sqrt{z_N}$. Therefore, 
such a direction is a good candidate for the inflaton.
Note that, when expressed in terms of the canonically normalized field, 
the large eigenvalue $z_N$ naturally suppresses {\it all} the shift-symmetry
breaking terms of $\Phi_{N}$, and the inflaton potential remains relatively 
flat  for $|{\rm Im}[\Phi_{N}]| \lesssim \sqrt{z_N}$.
The inflaton field excursion depends on the tensor-to-scalar ratio $r$, and
one needs typically $z_N$ greater than several tens for $r \gtrsim 0.01-0.1$.
Thus, large eigenvalues of $Z$  help us to have a good control of the inflaton potential over super-Planckian
field ranges.

The inflaton potential is lifted by shift-symmetry breaking terms. In order to account for the observed density
perturbations, the typical mass scale of the inflaton is of order $\GEV{12-13}$ for $r = {\cal O}( 0.01-0.1)$.
We assume that such inflaton potential arises from the following non-renormalizable interaction in the superpotetnial,
\beq
W &=& \kappa_{ij} \la QQ \ra X_i (\phi_j + \cdots) = m_{ij} X_i  (\phi_j + \cdots)
\label{WQQ}
\eeq
where $\la QQ \ra$ schematically represents a condensate of hidden quarks, and 
we have introduced a set of chiral superfields $X_i$~$(i=1-N)$ with an $R$-charge of $2$ and a negative
parity under $Z_2$. We have defined the mass matrix 
$m_{ij} \equiv \kappa_{ij} \la QQ\ra$, and  the required inflaton mass scale can be realized for e.g. 
$\la QQ \ra \sim (\GEV{16})^2$ and $\kappa_{ij} \sim 1$.  Because of the $R$-charge assignment, $\{\phi_i\}$ forms a Dirac mass $m_{ij}$  with $\{X_j\}$. The dots represent higher order terms in $\phi_i$ such as $\phi_i \phi_j \phi_k$ etc. We focus on the leading term in $\phi_i$ below to see the qualitative features.
The effect of higher order terms will be studied in the next section.

Let us rewrite the superpotential as
\bea
W &=&   M_{ij} X_i \Phi_j,
\eea
where we  have defined $M_{ij} \equiv  m_{ij} R_{jk}/\sqrt{z_{k}}$.  In the following we assume $\{X_i\}$ is canonically
normalized. 
We are interested in a case where ${\rm Im}[\Phi_N]$ takes super-Planckian field values. Let us rotate $X = V {\tilde X}$
by a unitary matrix $V$ so that $\tilde M \equiv  V^T M$ satisfy $\tilde M_{ij} = 0$ for $i>j$. Focusing on the ${\tilde X}_N$, the superpotential looks like
\beq
W = \tilde X_N \left( \tilde M_{NN} \Phi_N + \sum_{I=1}^{N-1} \tilde M_{NI} \Phi_I \right) + W_H(\tilde X_I, \Phi_I),
\eeq
where $W_H$ does not depend on $\tilde X_N$ or $\Phi_N$, and  here and in what follows the 
subscript $I$ runs from $1$ to $N-1$. 
When ${\rm Im}[\Phi_N]$ develops a super-Planckian field value, the first term in the parenthesis gives a large contribution
to the $F$-term of $\tilde X_N$. For successful inflation, the $F$-term should not be absorbed by the other terms.
This can be realized in either of the following two ways. First, if $\Phi_I$ has a mass much heavier than
$|\tilde M_{NI}|$ from its interactions with $\tilde X_I$ in $W_H$, the $\Phi_I$ is stabilized by its large mass, and it cannot absorb  the $F$-term. Secondly,
if $|\tilde M_{NI}| < |\tilde M_{NN}|$, the $\Phi_I$ cannot absorb the $F$-term for $|\Phi_I| \lesssim 1$. Thus, the $F$-term
of $\tilde X_N$ receives its main contribution from $\la \Phi_N \ra$, if the other scalars $\Phi_I$ are either heavier
than $|\tilde M_{NI}|$ or their couplings to $\tilde X_N$ is sufficiently weak. 
This is tantamount to saying that $\tilde X_N$ and $\Phi_{N}$ form (more or less) degenerate mass eigenstates.

Now let us suppose that the universe is dominated by the $F$-term of $\tilde X_N$ during inflation.
%
Then the scalars $\Phi_I$ and $X_I$ heavier than the Hubble parameter during inflation can be integrated out.
The other ones including $\tilde X_N$  generically acquire the Hubble-induced mass 
from the Planck-suppressed interactions, $K \supset -|\Phi_I|^2|\tilde X_N|^2$ and $-|\tilde X_i|^2 |\tilde X_N|^2$.
 We therefore assume
that all the scalars other than the inflaton are stabilized at the origin during inflation.\footnote{
Precisely speaking, $\Phi_I$ develops the inflaton-dependent VEV, and there are also small mixings 
between $\Phi_N$ and $\Phi_I$. Also, if some of them acquire a negative Hubble-induced mass, 
they will be stabilized below the Planck scale and acquire a mass of order the Hubble parameter. 
Such deviation from the origin or the small mixings with the  $\Phi_N$ only result in 
slight modification of the effective inflaton potential, and the inflaton dynamics as well as the 
predicted $n_s$ and $r$ are not significantly modified.
} 
If these conditions are satisfied, the inflaton potential
is approximately given by~~\cite{Kawasaki:2000yn,Kawasaki:2000ws}
\beq
V_{\rm inf}(\varPhi_{N}) \approx \frac{1}{2} M^2 \varphi^2,
\eeq
where we have defined $\varphi \equiv \sqrt{2} \,{\rm Im}[\Phi_{N}]$ and $M \equiv |\tilde M_{NN}|$.
The above inflaton potential is valid up to $\varphi \sim \sqrt{z_N}$. 
Thus,  a simple quadratic inflation  can be realized because of
 the approximate shift symmetry with the aid of the kinetic term anarchy. Note that, in general, the
 inflaton condensate after inflation is composed of various mass eigenstates, because the mass eigenstates
 change in the course of the evolution of the inflaton. This could affect the inflaton decay process and
 thermal history after inflation.

So far we have simply assumed that the kinetic term coefficient matrix $Z_{ij}$ is subject to a certain random
distribution so that one (or more) of the eigenvalues can happen to be much larger than unity. 
Let us study a concrete example if this can be indeed realized. 
In order to ensure the positive definite eigenvalues, we assume that $Z_{ij}$ takes the following form,
\bea
Z &=& \frac{1}{N} Y^\dag Y,
\label{ZYY}
\eea
where $Y_{ij}$ is a complex-valued $N \times N$ non-singular matrix. If each element $Y_{ij}$ follows a complex 
normal distribution with the variance $\sigma=1$, the averaged eigenvalue densities  of $Z$ 
is given by the Marcenko-Pastur law~\cite{Marcenko1967},
\beq
f(z) = \frac{1}{2\pi} \sqrt{\frac{4-z}{z}}~~~~(z > 0),
\eeq
in the large $N$ limit, and the averaged eigenvalue is unity.  For a finite value of $N \gg 1$, the averaged  largest
eigenvalue $z_N$ is given by $\la z_N \ra \approx 4$, and its typical fluctuations is of 
order $N^{-2/3}$~\cite{Johnstone2001,Johansson}.
Therefore, the typical value of $z_N$ is not large enough to realize the large-field inflation 
with $r = {\cal O}(0.01-0.1)$.\footnote{
The situation does not improve even if one considers a rectangular matrix for $Y_{ij}$.
}
The probability for $z_N$ to take atypical large value  is  exponentially 
suppressed, but it is non-zero~\cite{Majumdar:2009zza}:
\beq
P(z_N) \sim \exp\left[-  N z_N \right]~~~{\rm for}~~ z_N \gg 4.
\eeq
If the kinetic term coefficient matrix $Z_{ij}$ is determined by the vacuum expectation values (VEVs) of the
moduli fields, and if there are a sufficiently large number of different vacua in the string landscape, 
  it may be possible to realize $Z_{ij}$ with $z_N$ greater than several tens,
which support large-field inflation. Such selection of $Z_{ij}$ may be justified by anthropic arguments.

We may consider another distribution for $Z_{ij}$. Indeed,
there is no a priori reason to assume that $Z_{ij}$ is given by the square of random matrices
like (\ref{ZYY}). If the random nature of $Z_{ij}$ reflects complicated moduli stabilization dynamics,
 it is not unlikely that it involves some non-trivial dependence on the random matrices. 
 For instance it will be more likely to realize $z_N \gg 1$ if the distribution is proportional
 to $\exp[Y^\dag Y/N]$.  Alternatively, it is also possible 
 that $Z$ involves the inverse of the random matrices. In either case,  the largest eigenvalue $z_N$ can be easily much larger than unity, even if the typical eigenvalues are of order unity.

The number of singlets is also likely to have an important effect on how large $z_N$ can be, 
although it is difficult to estimate it quantitatively because of our ignorance of the UV theory.  
Our main point is that, if $Z_{ij}$ is subject to a certain random distribution, 
the kinetic term coefficient for one of the singlets could happen to be enhanced,
which helps to realize a flat potential over super-Planckian field ranges suitable for large-field inflation.

\subsection{Emergent shift symmetry}
Now we go on to the second model. In the previous case we have assumed the shift symmetry for
the quadratic term in the K\"ahler potential. Here we discuss a possibility that such  shift symmetry
arises from the anarchic nature of both holomorphic and non-holomorphic quadratic couplings of the
singlets.

First of all, let us note that, in global supersymmetry, the holomorphic quadratic terms
do not contribute to the scalar potential, and therefore, the enhancement of the kinetic term
coefficient suffices for the flat potential over super-Planckian ranges. 
In supergravity, on the other hand, the existence of
the holomorphic terms in the K\"ahler potential is crucial for the inflaton to take super-Planckian field values. 
This is because of the exponential factor in the supergravity scalar potential; the inflaton component should
not appear in the K\"ahler potential with unsuppressed coefficients. 

Let us consider the K\"ahler potential for the singlet scalars $\{\phi_i\} \,\,(i=1-N)$, which 
generically contain both holomorphic and non-holomorphic quadratic couplings like $\phi_i \phi_j + {\rm h.c.}$
and $\phi_i^\dag \phi_j$.  Here we rearrange the K\"ahler potential in such a way that anarchic nature of those
two kinds of coupling coefficients can be treated in a unified manner:
\bea
K\;=\; \frac{1}{2} Z_{ab} \varphi_a \varphi_b + \cdots,
\label{em}
\eea
where $Z_{ab}$ is a real symmetric $2N \times 2N$ matrix,  and
we have defined
\bea
\varphi_a \equiv \left\{
\bear{cl}
-i(\phi_j-\phi_j^\dag) &{\rm ~~for~~} a=j\\
\phi_j+\phi_j^\dag &{\rm ~~for~~} a=N+j\\
\eear
\right.,
\eea
where $a$ and $j$ run from $1$  to $2N$ and $1$ to $N$, respectively. 
We assume that various Planck-suppressed operators  shown by the dots in (\ref{em}) 
are such that, {\it if the eigenvalues of $Z_{ab}$ were of order unity}, 
the scalar potential would remain relatively flat along each $\varphi_a$ up to the Planck scale
when the other scalars are set to be at the origin. 

Let us diagonalize $Z_{ab}$ as
\bea
\varphi_a &=& R \tilde \varphi_a \\
R^T Z R &=& {\rm diag}(z_a) \equiv D_Z
\eea
with $z_1\leq \cdots \leq z_{2N}$, where $R$ is a $2N \times 2N$ orthogonal matrix. 
Suppose that $Z_{ab}$ follows a certain random distribution and that 
the largest eigenvalue $z_{2N}$ happens to be much larger than unity,
while the others are of order unity.\footnote{
This condition sets a non-trivial relation between the holomorphic and non-holomorphic coupling
coefficients of the quadratic terms in the basis of $\{\phi_i\}$.
} Focusing on this direction, the K\"ahler potential 
\bea
K = \frac{z_{2N}}{2} \left(\tilde \phi_{2N} + \tilde \phi^\dag_{2N}\right)^2 + 
\frac{1}{2} \left(\tilde \phi_{2N} - \tilde \phi^\dag_{2N}\right)^2+
\cdots,
\eea
where we have defined $\varphi_{2N} \equiv \tilde \phi_{2N} + \tilde \phi_{2N}^\dag$
and dropped a numerical coefficient of order unity in the second term for simplicity,
and the dots represent the other quadratic terms as well as higher order terms with a coefficient of order unity. 
Note that the imaginary component of $\tilde \phi_{2N}$ is given by a certain combination of
$\phi_i+\phi_i^\dag$ and $\phi_i-\phi_i^\dag$.
The above K\"ahler potential exhibits an approximate shift symmetry along the imaginary
component of $\tilde \phi_{2N}$.

The kinetic term coefficient of both the real and imaginary components of $\tilde \phi_{2N}$
is roughly given by $z_{2N} \gg 1$, which suppresses the reaction of these components to the
potential term and effectively enhances the friction on the motion. This would lead to the inflation
if the potential remains relatively flat up to the Planck scale. However, the scalar potential
for the real component of $\tilde \phi_{2N}$ becomes exponentially steep at sub-Planckian
values because of the enhanced kinetic term coefficient. On the other hand, the imaginary component,
${\rm Im}[\tilde \phi_{2N}]$, can take a value close to the Planck scale without exponential steepening, 
because, even though
their kinetic term coefficient is large, they do not appear in the K\"ahler potential with an enhanced
coupling. That is why the chaotic inflation in supergravity is possible in the presence of a large
kinetic term coefficient with the (approximate) shift symmetry. 

One can see this in a more conventional language by making $\tilde \phi_{2N}$ canonically normalized:
\beq
\Phi_{2N} \equiv \sqrt{z_{2N}} \tilde \phi_{2N}.
\eeq
Then, the imaginary component of $\Phi_{2N}$ can take a super-Planckian field value up to
$|{\rm Im}[\Phi_{2N}]| \lesssim \sqrt{z_{2N}}$, which realizes the chaotic inflation.

\section{Polynomial chaotic inflation}

Now let us consider concrete models. If the shift symmetry is accidental (or emergent), there are
generically various shift-symmetry breaking terms, which can modify the inflaton potential from the 
simple quadratic potential.  Therefore, we expect that the inflaton potential can be approximated 
by polynomials in the inflaton.
We revisit the polynomial chaotic inflation in supergravity proposed by 
the present authors as such~\cite{Nakayama:2013jka,Nakayama:2013txa}.\footnote{The polynomial chaotic inflation in supergravity was also studied 
 in Refs.~\cite{Kobayashi:2014jga,Kallosh:2014xwa} after BICEP2.}
The  K\"ahler and superpotential for $\phi_i$ and $X_i~(i=1-N)$ are
\beq
\begin{split}
	K  &= \frac{1}{2}Z_{ij}(\phi_i + \phi_i^\dagger)(\phi_j + \phi_j^\dagger) +|X_i|^2 +  K_{NR}, \\
	W &= X_i \left(m_{ij} \phi_{ij} +  \lambda_{ijk\ell} \phi_j \phi_k  \phi_\ell+ \cdots \right)
\end{split}
\eeq
where $\phi_i$ respects the shift symmetry on the quadratic terms, and
$K_{NR}$ denotes higher order terms in the K\"ahler potential, 
which include terms like $X_i X_j X_k^\dag X_\ell^\dag$ as well as 
shift-symmetry breaking terms such as $\phi_i\phi_j \phi_k \phi_\ell$, $\phi_i\phi_j \phi_k^\dagger \phi_\ell^\dagger$, etc. with coefficients of $O(1)$. We assume that $m_{ij}$ and $\lambda_{ijk\ell}$ are the same order 
in the Planck unit (cf. \EQ{WQQ}) and higher-order terms with coefficients of the same order are represented by the dots.
We have imposed $Z_2$ symmetry under which $\phi_i \to -\phi_i$, $X \to -X$ and assigned U(1)$_R$ charges
as $R(\phi_i)=0$ and $R(X_i)=2$. 

After diagonalizing the kinetic term by using $ \Phi = \sqrt{D_z} R^T \phi$, we obtain
\beq
\begin{split}
	K  &=\frac{1}{2} (\Phi_i+\Phi_i^\dagger)^2 + |X|^2+ K_{NR},\\
	W & = X_i  \left(\tilde m_{ij} \Phi_i +  \tilde \lambda_{ijk\ell} \Phi_j \Phi_k \Phi_\ell+ \cdots \right),
\end{split}
\eeq
where $\tilde m_{ij} = m_{ik}R_{kj}/\sqrt{z_j}$ and $\tilde \lambda_{ijk\ell}= 
\lambda_{iabc} R_{aj}R_{bk}R_{c\ell}/
\sqrt{z_j z_k z_\ell}$.
As described in Sec.~\ref{sec:kin}, the imaginary component of  $\Phi_N$ 
can be regarded as the inflaton, because the scalar potential does not exponentially blow up
for  $ |{\rm Im}[\Phi_N]| \lesssim \sqrt{z_N}$. The relevant terms including $\Phi_N$ can be rewritten
as
\bea
W_{\rm eff} \simeq  X^{(1)}_N  M_{NN} \Phi_N +  X^{(2)}_N
\tilde \lambda_{NNNN} \Phi_N^3 + \cdots,
\eea
where we have defined $X^{(1)} = U X$ and $X^{(2)}=V X$ for some unitary matrices $U$ and $V$
such that $U_{ji} \tilde m_{jN} = 0$ and $V_{ji} \tilde \lambda_{j NNN}=0$ for $i < N$. We assume that 
the other fields are either sufficiently heavy or their couplings to $ X^{(1)}$ and $X^{(2)}$ are sufficiently
weak so that the $F$-terms of $X^{(1)}_N$ and $X^{(2)}_N$ mainly come from the VEV of $\Phi_N$. 
If $X^{(2)}_N$ can be expressed in terms of only $X^{(1)}_I$ $(I=1-N-1)$, 
the $F$-term potentials of $X^{(1)}_N$ and $X^{(2)}_N$ induce the quadratic and hexic potentials, respectively.
In general, $X^{(1)}_N$ and $X^{(2)}_N$ are not independent from each other, and their overlap generates 
the quartic potential as a cross term. Therefore, the resultant inflaton potential generically takes the polynomial 
form.\footnote{Even if the other scalars are not stabilized at the origin but somewhere below the Planck scale,
the effective inflaton potential can be well approximated by the polynomial potential with
slightly different numerical coefficients.}

Let us consider a limiting case of $X^{(1)}_N = X^{(2)}_N$.
Then, the relevant K\"ahler and super-potentials are given by
\beq
\begin{split}
	K  &= \sum_i \frac{1}{2}(\Phi_i+\Phi_i^\dagger)^2 + |X|^2+ K_{NR},\\
	W & = X \left( M \Phi + \lambda \Phi^3 + \dots \right),
\end{split}
\eeq
where we have defined $X= X^{(1)}_N$, $\Phi = \Phi_N$, $M \equiv M_{NN}$ 
and $\lambda \equiv\lambda_{NNNN}$, and we expect $|\lambda| \sim  |M|/z_N$ if $m_{ij} \sim \lambda_{ijk\ell}$
(see \EQ{WQQ}).
This is exactly the same form as the polynomial 
chaotic inflation proposed  in Refs.~\cite{Nakayama:2013jka,Nakayama:2013txa}. 
The inflaton potential is given by
\begin{equation}
	V \simeq \frac{1}{2} |M|^2\varphi^2 \left( 1-\frac{|\lambda| \cos\theta}{|M|}\varphi^2
	 + \frac{|\lambda|^2}{4|M|^2}\varphi^4 \right),
\end{equation}
where we have defined $\varphi \equiv {\rm Im}[\Phi]$ and $\theta \equiv {\arg}[\lambda/M]$.
Because of the largeness of $z_N$, higher order shift symmetry breaking terms in the K\"ahler potential are
 suppressed which enables successful inflation.
For example, the terms quartic  in the  inflaton like $K_{NR} \propto |\Phi|^4$ are accompanied with
a suppression factor $\sim 1/z_N^2$, which induces a correction to the inflaton potential of order
$H_{\rm inf}^2 \varphi^4/z_N^2$. The correction to the inflaton mass remains sufficiently small
for $\varphi \lesssim \sqrt{z_N}$.

The potential is naturally given by the polynomial and its starts to deviate from the quadratic
 potential at $\varphi \sim \sqrt{|M/\lambda|} \sim \sqrt{z_N}$, which may be close to the point above which
 various other corrections become important. If $ M_{NN}$ is slightly smaller than $ \lambda_{NNNN}$,
there will be a field range where the inflaton potential is well approximated by the above polynomial inflation. 
 The prediction on $n_s$ and $r$ can thus be significantly deviated from those of the chaotic inflation with
a monomial potential. Importantly,  as emphasized in Refs.~\cite{Nakayama:2013jka,Nakayama:2013txa},
they cover the entire range allowed by the Planck results.

\section{Discussion and Conclusions}

So far we have imposed the $Z_2$ symmetry on the scalars to simplify the analysis. In fact, such $Z_2$ symmetry
suppresses non-thermal gravitino production from the inflaton decay~\cite{Kawasaki:2006gs,Kawasaki:2006hm,
 Asaka:2006bv,Dine:2006ii,Endo:2007ih,Endo:2007sz}, avoiding tight  cosmological constraints~\cite{Nakayama:2014xca}. If the $Z_2$ symmetry remains unbroken, we need to assign the $Z_2$ charge on the standard model
 particles and their superpartners for successful reheating.
Interestingly, the $Z_2$ symmetry can be identified with the matter parity, which forbids baryon- and lepton-number violating operators dangerous for too fast proton decays.
Then the singlet scalars  play the role of  right-handed sneutrinos~\cite{Murayama:2014saa}. 
In particular, if we apply the kinetic anarchy  to the right-handed
sneutrino inflation, the shift symmetry may arise naturally~\cite{MKTY}. Let us briefly discuss this case below.

Let us consider the model (\ref{K1}) -(\ref{Phi}). Instead of introducing $\{X_i\}$, we consider
the following superpotential  for $\{\phi_i\}$:
\bea
W\;=\; \frac{1}{2} m_{ij} \phi_i \phi_j,
\eea
where  we do not impose U(1)$_R$ symmetry. The $\tilde \phi_N$ tends to be the lightest one because of the large
kinetic term coefficient.  Then, after integrating out heavy scalars, the lightest one forms either a Majorana mass
with itself or a Dirac mass with a partner, which we denote $X$. The successful inflation is possible in the latter case.
Interestingly, this scenario provides a unified picture of $\Phi$ and $X$ in the original model (\ref{WKYY}), because
 both  are originated from the singlet scalars $\{\phi_i\}$, and it is simply the eigenvalue
 $z_N \gg 1$ that makes the shift symmetry of $\Phi$ of better quality. The fact that only $\Phi$ has
 a shift symmetry of better quality is crucial for successful inflation, because,
if the shift symmetry on $X$ is a good symmetry,  the inflation path is destabilized by the appearance 
of a tachyonic direction. 

Another interesting aspect of this model is that the inflaton as well as other scalars have 
neutrino yukawa couplings  like
\beq
	W = y_{\alpha i}\phi_i L_\alpha H_u = Y_{\alpha i} \Phi_i L_\alpha H_u,  \label{yukawa}
\eeq
where $H_u$ is the up-type Higgs doublet and $L_\alpha$ denotes the lepton doublet with $\alpha=1,2,3$
being the generation index and $Y_{\alpha i} = y_{\alpha j} R_{ji}/\sqrt{z_i}$.
Since the inflaton mass scale is $\sim 10^{13}\,{\rm GeV}$, it explains the tiny neutrino masses observed by
neutrino oscillation experiments through the seesaw 
mechanism~\cite{Minkowski:1977sc,Yanagida:1979as,Ramond1979,Glashow:1979nm} for $y_\alpha \sim O(0.1)$.\footnote{
Note that the yukawa couplings (\ref{yukawa}) stabilize $L_\alpha$ and $H_u$ during inflation at the origin,
since both $X$ and $\Phi$ have large yukawa couplings.	In Ref.~\cite{Nakayama:2013txa}, it was assumed 
that only either $X$ or $\Phi$ has a renormalizable yukawa coupling	$y$ and derived a constraint as
 $y \lesssim 10^{-5}$ for successful inflation. This constraint can avoided in the present model because 
 both $X$ and $\Phi$ have yukawa couplings.
 }  The observed large neutrino mixings may be (partly) originated from the anarchic nature of the singlet
 scalars including the inflaton, which nicely fits with the neutrino mass anarchy hypothesis~\cite{Hall:1999sn,Haba:2000be}.

In this letter we have proposed a scenario in which an approximate shift symmetry required for large-field inflation
appears because of the enhanced kinetic term coefficient. Such an enhancement may be realized if there are
many singlets whose kinetic term coefficients follow some random distribution. Although the probability for such 
an enhancement to take place is exponentially suppressed in the simplest model, such fine-tuning may be
justified by the anthropic arguments. Alternatively, if the kinetic term coefficients follow a different distribution such 
as the exponential or inverse of the random matrix, it becomes much easier to obtain the required enhancement. In supergravity,
such an enhancement is not sufficient for successful large-field inflation, and a certain relation is required between
the holomorphic and non-holomorphic quadratic terms of the inflaton. We have studied two cases in which such a relation is satisfied because of the approximate shift symmetry or the relation emerges by accident. In our scenario,
the shift symmetry is only approximate, and so, the inflaton potential generically receives various shift-symmetry
breaking terms. We have shown that the resultant inflaton dynamics is well approximated 
by the polynomial chaotic inflation. Interestingly, the predicted 
values of $n_s$ and $r$ can be significantly deviated from those of the simple quadratic chaotic inflation. In fact,
it was shown in Refs.~\cite{Nakayama:2013jka,Nakayama:2013txa} that the predicted $(n_s, r)$ of the polynomial chaotic inflation can cover the entire ranges allowed by the Planck data. The future B-mode experiments, therefore, will be able to refute or support our inflationary scenario.

\section*{Acknowledgments}
We thank Hitoshi Murayama for fruitful discussion. TTY learned that Ken-iti Izawa is working on 
a similar subject.
This work was supported
by the Grant-in-Aid for Young Scientists (B) (No. 26800121 [KN], No.24740135  [FT]),   Scientific Research on Innovative Areas (No.23104008 [FT]), and Scientific Research (B) (No.26287039 [FT and TTY]), 
by Inoue Foundation for Science [FT], and by World Premier International Center Initiative (WPI Program), 
MEXT, Japan.

\bibliography{references}
%

\end{document}